\documentclass[twocolumn,superscriptaddress ]{revtex4-1}

\usepackage{graphicx}
\usepackage{hyperref}
\usepackage{amsmath,amssymb}
\usepackage{epstopdf}
\usepackage{subcaption}
\graphicspath{{figs/}}
\begin{document}

\title{Statistical Investigation of Micro-Avalanches of Three Dimensional Small-World Networks and their Boundary and Bulk Cross-Sections}

\author{M. N. Najafi}
\affiliation{Department of Physics, University of Mohaghegh Ardabili, P.O. Box 179, Ardabil, Iran}
\email{morteza.nattagh@gmail.com}
\author{H. Dashti-Naserabadi}
\affiliation{Physics and Accelerators Research School, NSTRI, AEOI 11365-3486, Tehran,Iran}
\email{h.dashti82@gmail.com}

\date{\today}

\begin{abstract}
In many situations we are interested in the propagation of energy in some small portions of a three dimensional system with dilute long-range links. In this paper sandpile model is defined on the three-dimensional small world network with real dissipative boundaries and the energy propagation is studied in three-dimensional system as well as the two-dimensional cross sections. Two types of cross section are defined in the system, one in the bulk and another in the system boundary. The reason for considering the latter is to make clear how the statistics of the avalanches in the bulk cross section tend to the statistics of the avalanches in the system boundaries as the concentration of long range links ($\alpha$) increases. This trend is numerically shown to be power law in a manner described in the paper. Two regimes of $\alpha$ are considered in this work. For sufficiently small $\alpha$s the dominant behavior of the system is just like that of the regular BTW, whereas for the intermediate values the behavior is non-trivial with some exponents that are reported in the paper. It is shown that the spatial extent up to which the statistics is similar to the regular BTW model scales with $\alpha$ just like the dissipative BTW model with the dissipation factor (mass in the corresponding ghost model) $m^2\sim \alpha$ for the three-dimensional system as well as its two-dimensional cross-sections. It is also shown that some hyper-scaling relations are violated for large $\alpha$s.
\end{abstract}


\maketitle

\section{Introduction}

Self-organized criticality (SOC) as a commonly occurring phenomenon in nature and society is a very important notion which refers to the intrinsic tendency of a wide class of slowly driven (open) systems to evolve spontaneously to a non-equilibrium steady state characterized by long range correlations and power law scaling behavior. SOC can occur in regular (discrete or continuous) systems, as well as random systems. Some examples of regular (or nearly regular) systems are forest fire (as a discrete system) \cite{drossel1993exact}, water droplets \cite{plourde1993water}, earthquake \cite{chen1991self} and superconducting avalanches \cite{field1995superconducting} (as continuous systems). There are also interests on the notion of the SOC on the complex networks from both theoretical and experimental sides. Examples are biological evolution \cite{bak1993punctuated} and signal propagation in neural networks \cite{beggs2003neuronal}. The complex networks \cite{albert2002statistical,boccaletti2006complex} describe a wide domain of physical and other systems ranging from biological \cite{jeong2001lethality} and neural networks \cite{levina2007dynamical} to internet \cite{faloutsos1999power,albert1999internet,adamic2000power}, social \cite{liljeros2001web}, coauthors \cite{newman2001scientific}, citation \cite{redner1998popular} and wealth \cite{garlaschelli2008effects} networks. Theoretical examples of the SOC on the complex networks are the SOC on finite range random networks~\cite{najafi2014bak}, the SOC model for brain plasticity~\cite{de2006self}, the small-world sandpile models~\cite{hoore2013critical} and dissipative sandpiles on the small-world networks~\cite{bhaumik2013critical}. \\
Among the complex network models, the small world network model introduced by Watts \textit{et al}~\cite{watts1998collective} has the ability to interpolate between the regular lattices and random networks. These systems carry simultaneously the effect of both regular and random network statistics. It is simply defined by adding extra random long-range links to the regular lattice without modifying the configuration of the regular links which connect the neighboring nodes. These systems are therefore tuned by the $\alpha$ parameter which is the percent of the number of extra long range links per node in the system (to be defined later). The sandpile models (as the simplest prototype of SOC phenomena) on small network systems was considered by many people \cite{de2002self,chen2008long,bhaumik2013critical,hoore2013critical}. The main feature of these investigations is the observation of three regimes: small $\alpha$ regime at which the properties are compatible with the regular sandpile model, the intermediate regime and the large $\alpha$ regime in which the random network properties are dominant~\cite{bhaumik2013critical}. It was revealed also that each amount of $\alpha$ implies a spatial length scale at which a cross-over between different behaviors occur~\cite{hoore2013critical}. \\
A critical attention seems to be necessary to be paid to three-dimensions in which the real experiments are done, e.g. \cite{beggs2003neuronal}, which has poorly investigated in the literature. Furthermore the question that how does an avalanche spread in a two or three dimensional system should be addressed since it yields valuable information about the experimental results and the exponents. Consider for example the experiment by Beggs \textit{et al}~\cite{beggs2003neuronal} in which a two-dimensional lattice of multi-electrodes was embedded into the rat cortex which is a three dimensional system. Although the signal propagation in this system is three-dimensional, the spark detectors had been prepared in some limited part of the system involving a small fraction of nodes of the whole system. Therefore it seems reasonable that the investigations on this system should involve the tracing of the avalanches to observe how the data spread in the whole and a part of the system, i.e. measure how the information of the avalanche expansion is projected to some portion of the system, e.g. two dimensions. In the theoretical side also, it is interesting to measure how the information of a $d+1$ dimensional system is projected to $d$-dimensional system depending on the fact that the removed dimension is spatial~\cite{dashti2015statistical} or temporal \cite{najafi2016scale}. \\
In this paper we consider the energy propagation of Bak-Tang-Weisenfeld (BTW) sandpile model on the three dimensional small world network. We show that the quantity $\alpha$ measures how a typical bulk site \textit{is close to} the boundaries. The propagation of energy in two-dimensional sheets are also studied by defining two two-dimensional cross-sections; one in the bulk of the system and another on the boundary. We show that the statistics of the bulk sheets becomes closer and closer to the boundary one as $\alpha$ increases. This trend occurs in a power-law fashion with the exponents reported in the paper. Importantly it is observed that the spatial extent up to which the statistics is similar to the regular BTW model scales with $\alpha$ just like the dissipative BTW model with the dissipation factor (mass in the corresponding ghost model) $m^2\sim \alpha$ for both three-dimensional (original) system and two-dimensional cross-sections. Some hyper-scaling relations are also tested in terms of $\alpha$. \\
The paper has been organized as follows: in SEC.\ref{sec:ASM} we introduce the sandpile model on the small world network and its mapping to two dimensions. The numerical results for three dimensions are presented in SEC. \ref{sec:3D}. The results for two dimensions are reported in to sub-sections: in the subsection \ref{sec:2D-distribution} the distribution functions are analyzed, and the fractal dimensions are devoted to the subsection \ref{sec:2D-FDs}. We close the paper by a summary of our results in SEC.\ref{Conclusion0}.

\section{The BTW model on the small world networks}
\label{sec:ASM}

\begin{figure*}
\begin{subfigure}{0.25\textwidth}\includegraphics[width=\textwidth]{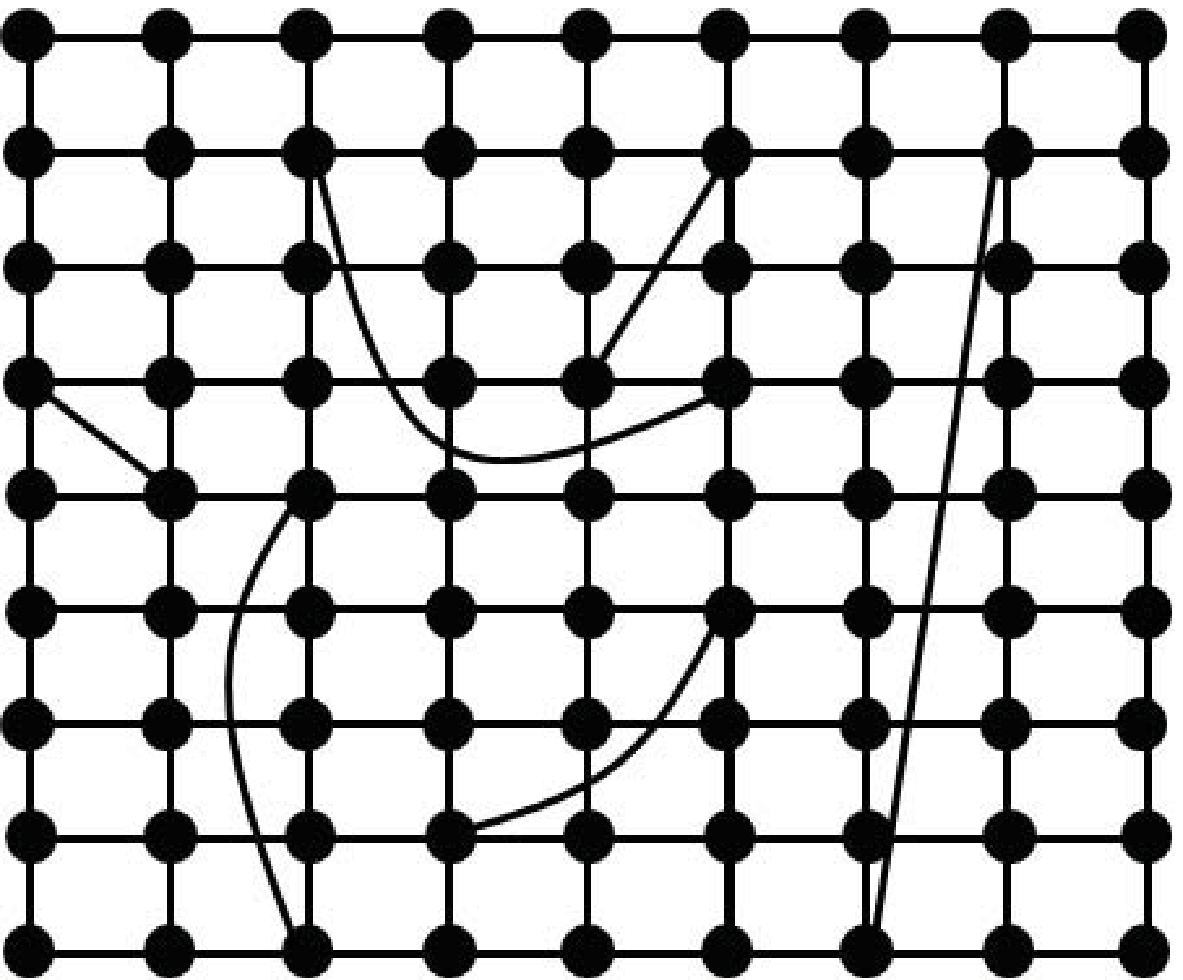}
\caption{}
\label{fig:2DLattice}
\end{subfigure}
\begin{subfigure}{0.45\textwidth}\includegraphics[width=\textwidth]{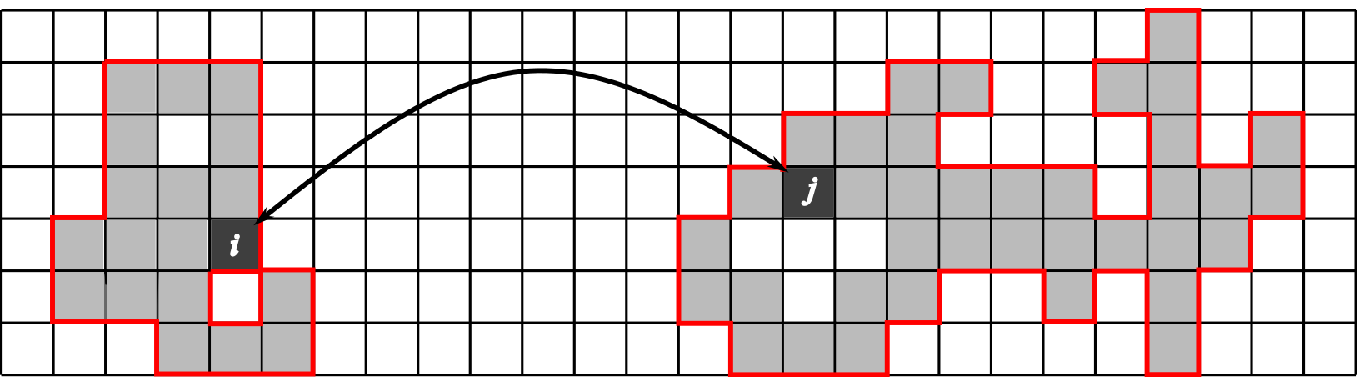}
\caption{}
\label{fig:Sink}
\end{subfigure}
\begin{subfigure}{0.4\textwidth}\includegraphics[width=\textwidth]{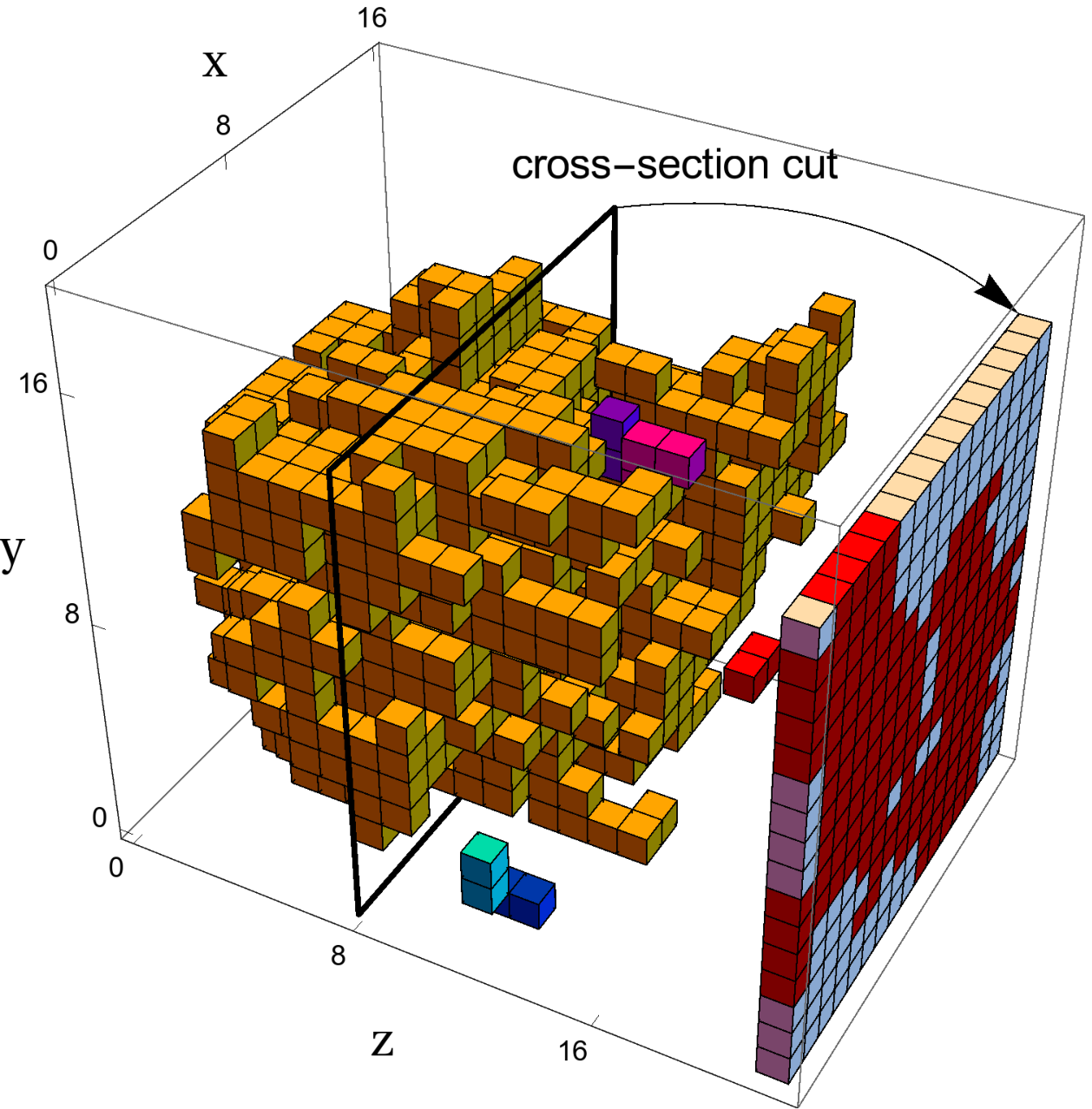}
	\caption{}
	\label{fig:3d-cross}
\end{subfigure}
\caption{(Color Online) (a) Schematic set up of a small world network with some random long-range links. Note that the regular bonds are not altered. (b) Two schematic micro-avalanches which are connected by a single link between two points $i$ and $j$.  The gray sites are the toppled sites and the red lines are the exterior frontiers of these micro-avalanches. Note that the point e.g. $i$ for the right hand micro-avalanche, plays the role of a \textit{sink point}. This is not however a perfect picture, since this point can play the role of the grain source as well. (c) The three-dimensional system with an avalanche and its projection in its cross section in $z_{\text{cross-section}}=L/2$.}
\label{fig:wave}
\end{figure*}
Let us first introduce the BTW model on the three dimensional cubic lattice with linear size $L$. The sand grains are distributed randomly throughout of the lattice, so that we have a local height filed $h$ over the lattice, for which the constraint is that no site has the height larger than $2d$, i.e. $h(i)$ takes the numbers from the set $\lbrace{1, 2, ... , 6}\rbrace$ for each site $i$. The system is open, i.e. adding or losing energy is allowed. The dynamic of the system is as follows: A random site ($i$) is chosen and a grain is added to this site, i.e. $h(i)\rightarrow h(i)+1$. If the resulting height is lower than a critical value ($h(i)<h_c=6$), another site is chosen for adding a grain. But if this height exceeds the critical value ($h(i)>h_c$), then this site becomes unstable and topples. During this toppling, the height of the original site $i$ is lowered by a number equal to its neighbors ($h(i)\rightarrow h(i)-6$) and the height of each of its neighbors increases by one in such a way that the total number of grains is conserved. The single toppling process can be expressed via the relation $h(i)\rightarrow h(i)-\Delta_{i,j}$ in which
\begin{equation}
\Delta_{i,j}=
\begin{cases} -1 & i \text{ and } j \text{ are neighbors} \\
6 & i=j \\ 
0 & \text{other}
\end{cases}
\end{equation}
As a result of this toppling, the neighboring sites may become unstable and topple. This process continues until reaching the state in which all sites of the system become stable. The chain of topplings is named as an avalanche. Now another site is chosen for injection and the process continues. Generally we have two kinds of configurations: transient and recurrent. The transient configurations may happen once in the early evolution steps and shall not happen again and the recurrent configurations take place in the steady state of the system. In this state, the energy input and output of the system is statistically equal and the statistical observables of the system are statistically constant. All of the configurations in this state occur with the same probability. For a good review see~\cite{Dhar1990Self}. The important aspect of this model is that the system organizes itself in the critical state. \\
The geometrical aspects of the pure two-dimensional regular BTW (which corresponds to $c=-2$ conformal field theory (CFT)) has been the subject of intense studies~\cite{Najafi2012Avalanche,Najafi2012Observation,Saberi2009Direct}. One example is the exterior perimeter of an avalanche which is numerically shown to be loop-erased random walk (LERW) in two dimensions~\cite{Najafi2012Avalanche,Saberi2009Direct}. The problem of exact enumeration of the critical exponents in 3D is more serious than 2D which has its roots in the rapid growing of the computational labor with the system size. In 3D the exterior perimeter of an avalanche is a fractal closed surface and is expected to scale with the toppled volume and its gyration radius.\\
Now let us consider the model on the cubic small world network. The dimensionality of such systems seems to play a crucial role and the exponents vary by the dimension for an $\alpha$. In Fig. \ref{fig:2DLattice} we have shown schematically a two-dimensional small world network (this figure has been sketched in two dimensions for simplicity). As is evident in this figure in addition to the regular links between neighbors, there are some long-range links between random-chosen sites. For this system we have two dependent further random fields in addition to the height field $h(i)$. The connection matrix $\text{L}(i,j)$ is unity if sites $i$ and $j$ (not neighbors) are connected by a long-range link and zero otherwise. The distribution of lengths and the degree of nodes are chosen to be uniform in the interval of allowed values (naturally the lengths are restricted to the linear size of the system). The other one is $z_c(i)=6+\sum_{j}\text{L}(i,j)$ which accounts the number of total links in the node $i$.
In this language if the height of a node exceeds $z_c(i)$ it topples according to the rule $h(i)\rightarrow h(i)-\Delta_{i,j}$ in which:
\begin{equation}
\Delta_{i,j}=
\begin{cases} -1 & i \text{ and } j \text{ are neighbors or $\text{L}(i,j)\neq 0$} \\
z_c(i) & i=j \\ 
0 & \text{other.}
\end{cases}
\end{equation}
$\alpha$ (the percent of long range links) is defined as follows \cite{hoore2013critical}:
\begin{equation}
\alpha\equiv 100\times\frac{1/2\sum_{i,j}\text{L}(i,j)}{\text{\# total regular links}}
\end{equation}
in which the factor $\frac{1}{2}$ is to prevent double counting (for Fig. \ref{fig:2DLattice} $\alpha=\frac{25}{6}$).  In this paper we are dealing with low values of $\alpha$, i.e. $\alpha=0.1,0.25,0.5,1,2,5$ and $10$. This model shows some common features with the ordinary regular BTW model. An example is that the states are classified into two categories: transient and recurrent states. Despite of many attempts, the local and global properties of this model are poorly understood especially in three dimensions. In this paper we consider and analyze the \textbf{connected components} of each avalanche and name them as \textit{micro-avalanche}. Let us describe it more: In a small world network, due to the existence of long range links, an avalanche may be composed of many connected components each of which is connected to the others by one or more long range links. Therefore the statistics of a single component is apparently different from the total avalanche as a whole. These micro-avalanches have their own mass, size, gyration radius (See Fig. 9 of reference \cite{bhaumik2013critical}). Two such micro-avalanches are schematically shown in Fig.\ref{fig:Sink} from which one can think of the connecting sites as the sink points. In this figure the sites $i$ and $j$ are connected by a long-range link and two connected components of an avalanche have been shown. An evident effect of long-rage links in these systems is the more direct effect of dissipative boundary sites to dynamics of grains. This is due to the fact that the effective distance of the bulk sites from the boundary sites (and any other bulk site) decreases as the concentration of long-range links increases. It can be understood more directly by the fact that the average shortest distance $\left\langle l\right\rangle $ is related to the system size $L$ and $\alpha$ via the relation $\left\langle l\right\rangle=LF\left(\alpha^{1/d}L\right)$ in which $d$ is the system dimension and $F(x)\propto \text{constant}$ for $x\ll 1$ and $F(x)\propto \frac{\log x}{x}$ for $x\gg 1$\cite{bhaumik2013critical}. Noting that for large enough $\alpha$s $F$ is a decreasing function, we see that the average shortest distance decreases with increasing $\alpha$. An important observation was recently made in two dimensions by Moghimi \textit{et al.}~\cite{hoore2013critical} in which it has been shown that for the small scales the system behaves like ordinary (regular) BTW model, whereas for the large scales the critical properties of the system changes crucially.\\
In the following sections we analyze global and local properties of this model in 3D and effective 2D systems. Our analysis for 3D avalanches involves the scaling relation between the global quantities and their distribution functions, as well as local ones. The three dimensional quantities studied in this paper are as follows:\\
- The avalanche mass $(M_3)$ which is the total number of sites involved in a three-dimensional micro-avalanche.\\
- The three-dimensional gyration radius $R_3$ which is defined as $ R_3^2\equiv \frac{1}{M_3}\sum_{i=1}^{M_3}\left|\vec{r}_i-\vec{r}_{\text{com}}\right|^2$, in which the sum runs over the points involved in a three-dimensional micro-avalanche. In this formula $\vec{r}_i\equiv(x_i,y_i,z_i)$ is the position vector of the $i$th point of the micro-avalanche and $\vec{r}_{\text{com}}\equiv \frac{1}{M_3}\sum_{i=1}^{M_3}\vec{r}_i$ is the center of mass of the micro-avalanche.\\
- The number of topplings in a three-dimensional micro-avalanche $(n_{\text{toppling}})$.

\subsection{Mapping to Two Dimensions}
\label{sec:map2d}

The problem of two-dimensional propagation of energy in three dimensional systems seems to be very important from both theoretical and experimental sides. The example mentioned in the previous section is the experiment by Beggs, in which the signal activity in the external cortex of rat was measured in an effective $60\times 60$ two-dimensional lattice which can be imagined as a system embedded in a three dimensional one. To get closer to the Beggs's experiment one may ask the same question for the small world networks and trace the avalanche dynamics of the cross sections of three dimensional system.\\
From the theoretical side, the important question is how the information in $d+1$ dimensions would be reflected to the $d$ dimensions. For this purpose one should map the original $d+1$ dimensional model to a $d$-dimensional one and measure how some information are lost and how the degrees of freedom in the subtracted dimension affect the $d$-dimensional model, i.e. find the model which lives in the lowered dimensional system. If the subtracted dimension be temporal, then one is looking at a \textit{frozen} model with no dynamics. The investigation of the contour lines of statistical systems~\cite{Najafi2012Observation} and ground state of the quantum systems~\cite{najafi2016scale} are some examples. A more interesting situation is the case in which the subtracted dimension is spatial one, like holographic principle. The other example is the cross sections of three-dimensional BTW model which is proposed to share some critical behaviors as the 2D Ising model~\cite{dashti2015statistical}. This investigation on the small world seems to be more interesting from the experimental side, for the reasons stated above. This motivated us to study the critical properties of the two-dimensional cross sections of the BTW model on the small world networks. The procedure of extracting the cross-section data from the three dimensional system has schematically shown in the Fig. \ref{fig:3d-cross}. \\
The three-dimensional and the effective two-dimensional energy propagation in small world systems is the aim of the present paper. The induced criticality of the resulting two-dimensional system is shown to be completely different from the three-dimensional case. The quantities which are analyzed in the cross-sections are the followings:\\
- The mass of 2D micro-avalanches $M_2$ which is the total number of sites involved in a 2D cross-section of a micro-avalanche.\\
- The loop lengths $l$ which is the length of the loop that is the external perimeter  of a 2D cross-section of a micro-avalanche. A loop sample corresponding to a 2D micro-avalanche is schematically shown in Fig. \ref{fig:Sink} in which the external frontier of the toppled region (micro-avalanche) has been extracted and shown by red lines. The loop length is simply the number (the total length) of these lines. \\
- The area inside loops $a$ which is the total area that is contained in the loop.\\ 
- The gyration radius of loops $r$ and 2D areas $R_2$ which have been defined above for the cross-sections.\\
- The number of topplings in an avalanche in the cross-section micro-avalanche $n_{\text{toppling}}$.\\

Two kind of cross section have been considered in this work, all of which are perpendicular to an axis (say $z$-axis): one is $z=\frac{L_z}{2}$ plate and the other $z=0$ plate in which $L_z$ is the linear size of the system in the $z$ direction. The last one is the most dissipative plate with the exponents completely different from the bulk cross-section, as we will see.

\section{Three dimensions}
\label{sec:3D}

In this section we present the numerical results for the three dimensional avalanches and 2D propagation of avalanches is postponed to SEC.~\ref{sec:2D}. We have considered $L\times L\times L$ cubic lattice with random long-range connections and with various sizes $L=50,100,200$ and $300$. The randomness can easily be established by choosing randomly pair of sites $i_0$ and $j_0$ in such a way that $\text{L}(i_0,j_0)=1$, a number of times corresponding to an $\alpha$ value. Therefore the long-range links in the resulting lattice have uniform length distribution and the degree distribution of nodes is apparently uniform. For all lattice sizes after some steps, the system reaches the steady state from which the samples have been extracted and the statistical analysis has been performed. In sandpile models to have nearly independent samples one can consider the time period of rare events as the time between two successive samplings. This rare event can be the event of very large avalanches. Let us define a \textit{large avalanche} as an avalanche which contains $n\geq (L_x L_y)/10$ sites inside. Our experience has been that the average time period for these rare events is $\left\langle \tau\right\rangle\approx 10$ (time step$\equiv$ the number of grain adding to the system). This can be interpreted as an event in which the height configuration is thoroughly updated and the memory is nearly lost. To be sure that this \textit{making independent} procedure had been more complete, we have let $100$ sand injections between two successive samplings, i.e. after extracting each sample, $10^2$ sand grains were randomly added to the system each of which causing a relaxation process, and then another sample was extracted.
The time of beginning of the recurrent sates has been obtained automatically. It was defined as the time above which the average height does not change with the external injection and is nearly constant.
For each 3D sample, we have extracted also the height configuration and the toppling statistics of the 2D cross section for further analysis in SEC.~\ref{sec:2D}. Over $5\times 10^6$ samples for each $\alpha$ and lattices size have been generated for analyzing the statistics of 3D problem and their 2D cross sections have been extracted. Various fractal dimensions and distribution functions have been calculated. We have two types of injection: the first type are the injections to the bulk sites as is customary in the sandpile models, whereas the second type are the injections to the boundary sites, say in the $z=0$ plane. The corresponding cross sections in both cases are perpendicular to the $z$ axis and are of two types: bulk cross section which is the cross section containing sites in the $z=L/2$ plane and the cross section containing boundary sites in the $z=0$ plane. As will be seen, this method helps us to define a measure for closeness of bulk sites to the boundary sites. In fact, as stated in the previous section, the larger the parameter $\alpha$ is, the closer the bulk sites to the boundary sites are. Therefore one expects that for larger values of $\alpha$ the statistics of the avalanches in the bulk sites is closer to the statistics of the avalanches whose injection points (first unstable point) are in the boundary sites. \\
The geometrical quantities of interest in this part of the paper are $x=M_3,R_3$ and $n_{\text{toppling}}$, defined in SEC.~\ref{sec:ASM}. The distribution functions of these quantities in the critical state are expected to behave like $N(x)\sim x^{-\tau_x}$ (in which $\tau_x$'s are their corresponding exponents), up to a specific scale above which the functions fall off more rapidly than the power-law behavior. Based on the above argument, it is obvious that for the larger values of $\alpha$ the extent of the power-law behavior is smaller, i.e. the finite size effects dominate earlier than smaller values of $\alpha$ since for larger values of $\alpha$ the typical shortest path from bulk sites to the boundary sites are smaller. We have also found that these quantities are related via the scaling relation $x\sim y^{\gamma_{xy}}$ in which $\gamma_{xy}$ are the scaling exponents as expected.\\
\begin{figure*}
\centering
\begin{subfigure}{0.45\textwidth}\includegraphics[width=\textwidth]{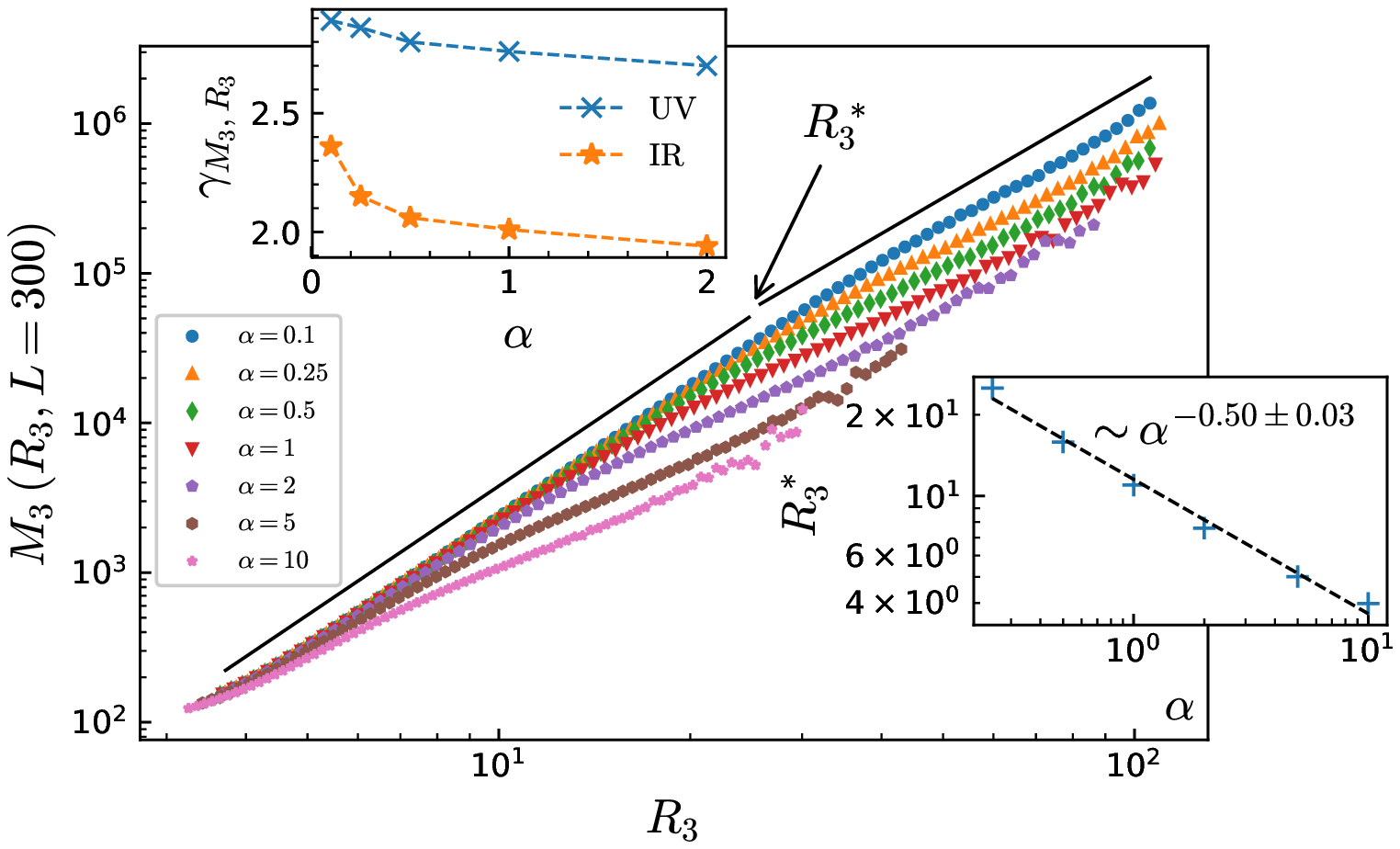}
\caption{}
\label{fig:M3_R3}
\end{subfigure}
\begin{subfigure}{0.45\textwidth}\includegraphics[width=\textwidth]{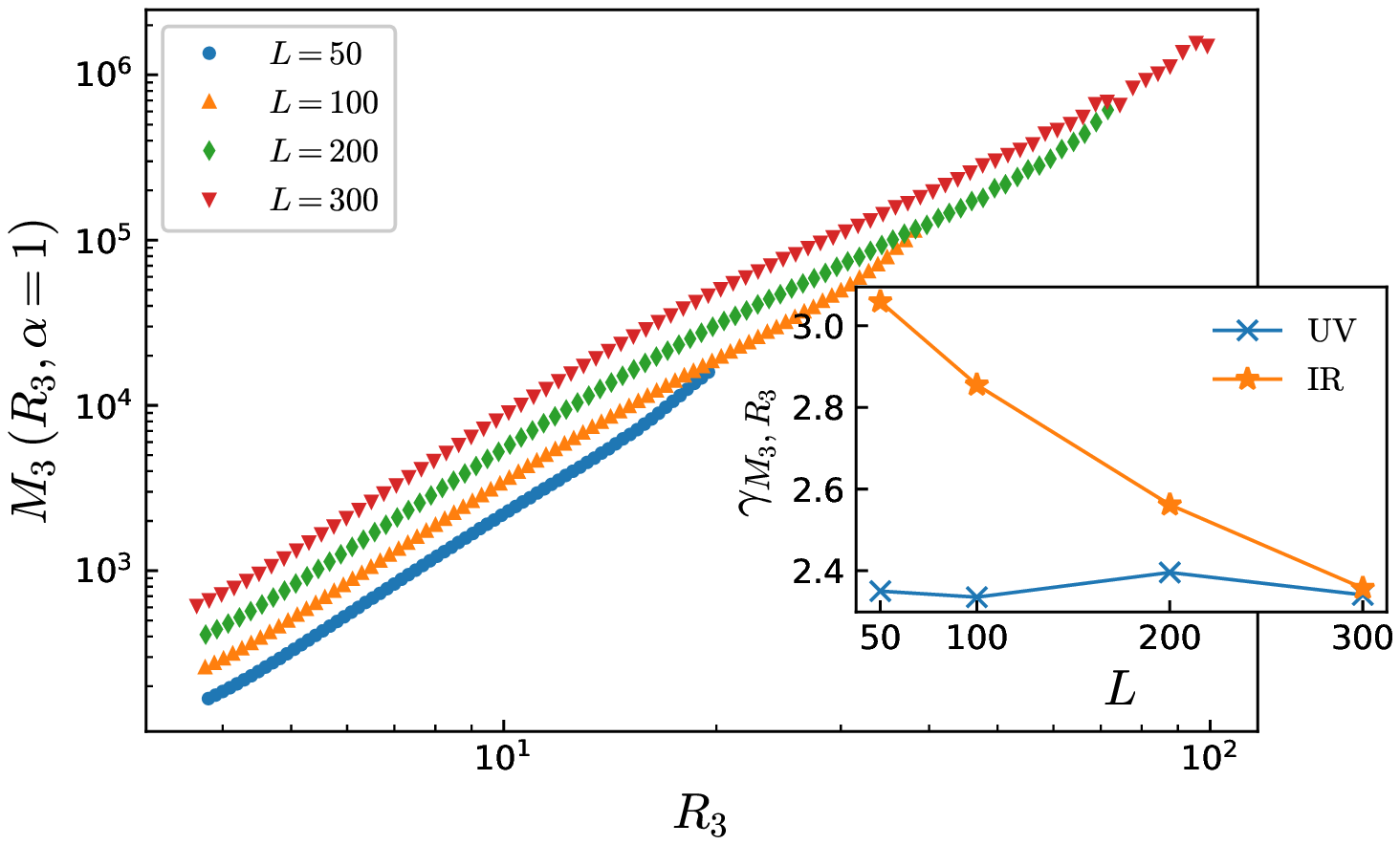}
	\caption{}
	\label{fig:M3_R3-alpha1}
\end{subfigure}
\begin{subfigure}{0.45\textwidth}\includegraphics[width=\textwidth]{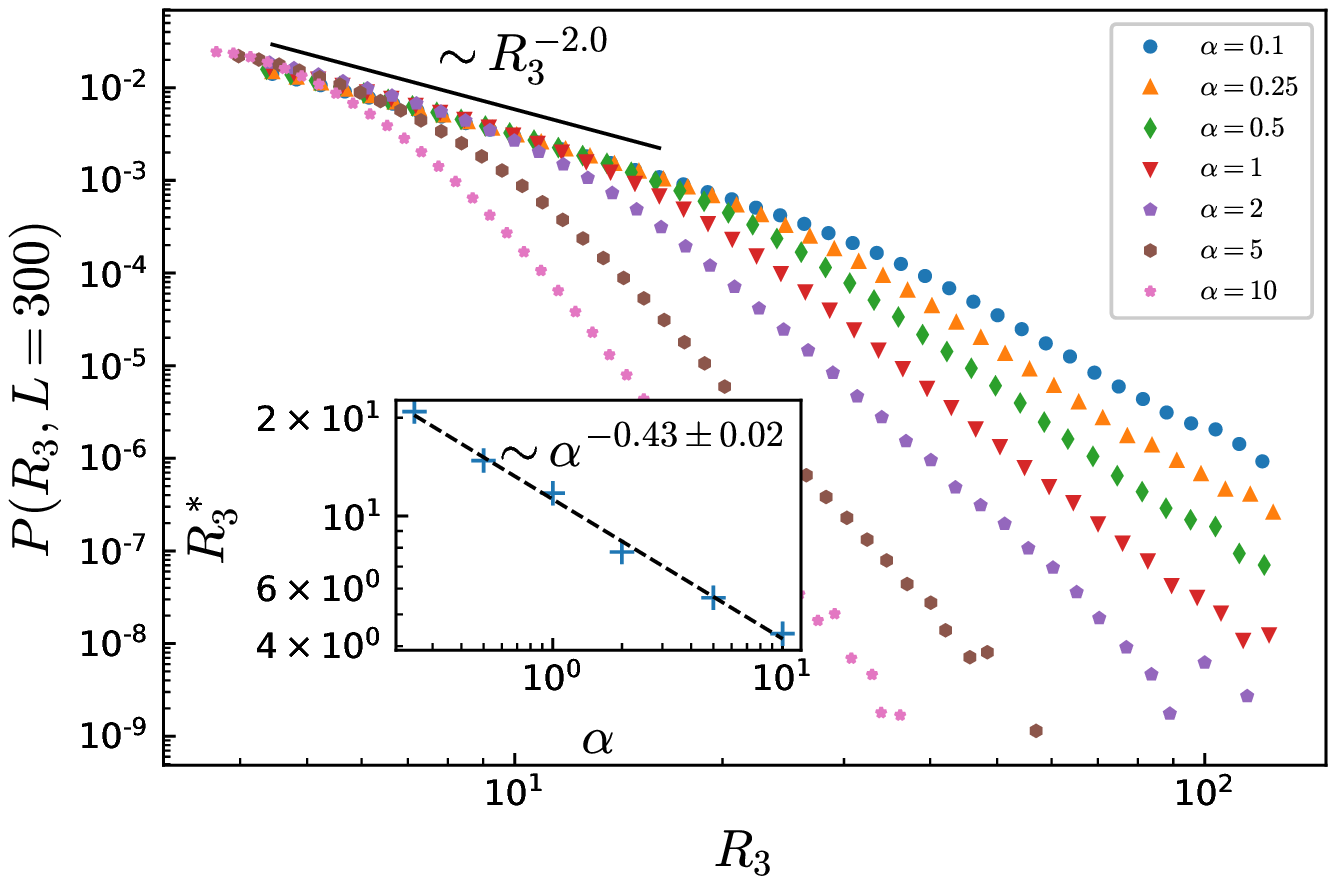}
\caption{}
\label{fig:P_R3}
\end{subfigure}
\begin{subfigure}{0.45\textwidth}\includegraphics[width=\textwidth]{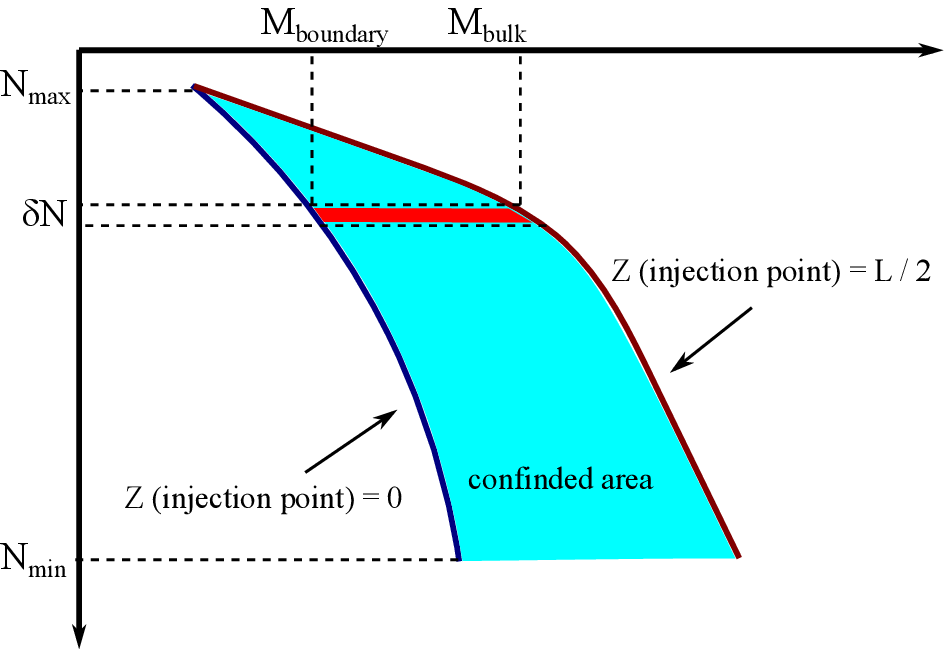}
\caption{}
\label{fig:confined}
\end{subfigure}
\begin{subfigure}{0.45\textwidth}\includegraphics[width=\textwidth]{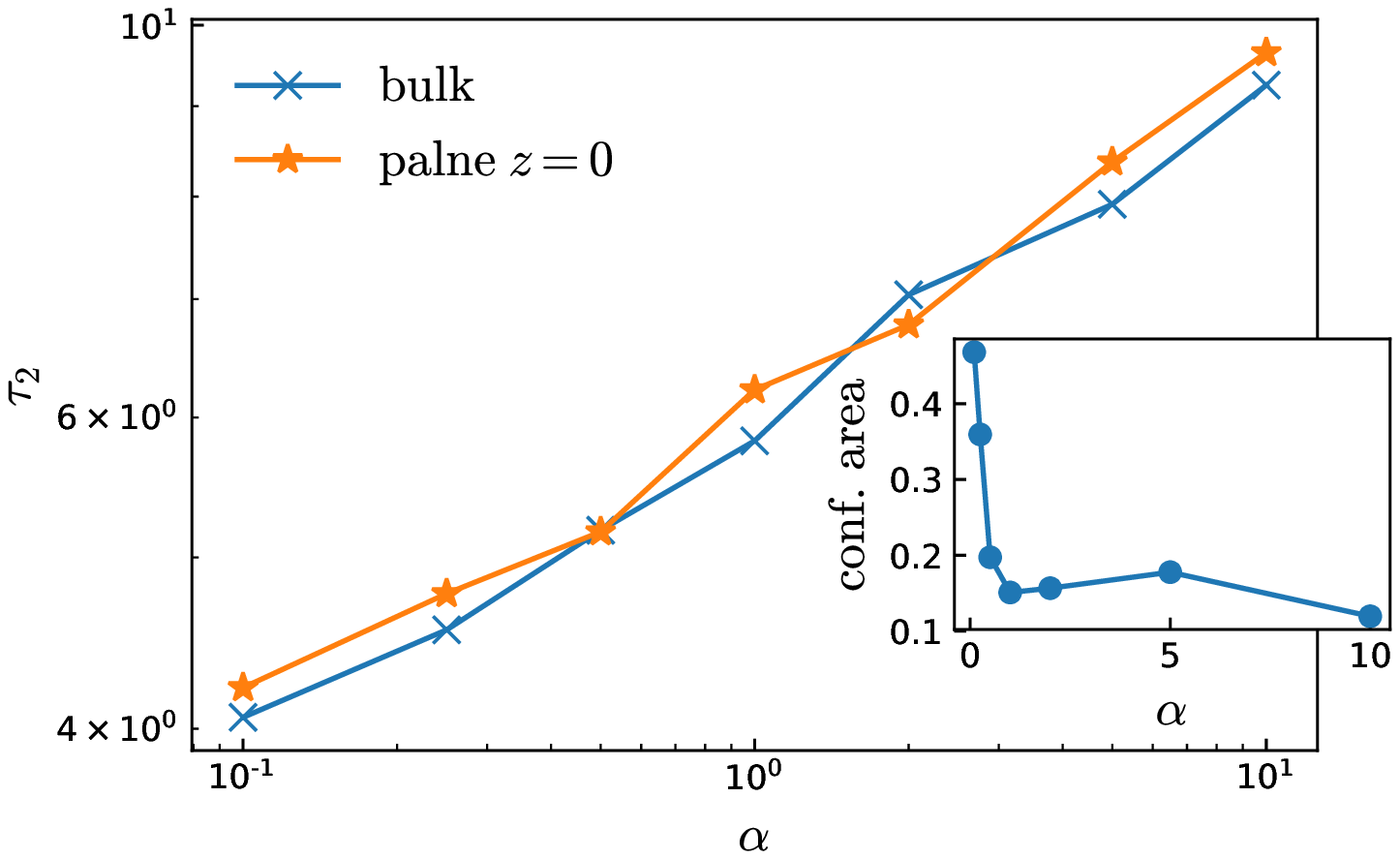}
\caption{}
\label{fig:tau2_alpha}
\end{subfigure}
\caption{(a) The plot of $M_3$ in terms of $R_3$ and the corresponding exponents $\gamma_{M_3R_3}$. Upper inset: $\gamma_{M_3R_3}^{\text{UV}}$ and $\gamma_{M_3R_3}^{\text{IR}}$. Lower inset: the cross-over radius $R_3^*$ in terms of $\alpha$ with the exponent $0.5\pm 0.03$. (b) The same as (a) for various lattice sizes $L$. The finite size dependent (UV and IR) slopes $\gamma_{M_3R_3}$ have been shown in the inset. (c) The log-log plot of the distribution function $P(R_3)$ in terms of $R_3$). Inset: the cross-over point $R_3^*$. (d) Schematic representation of the confined area between the distribution function2 of $M_3$ for bulk and boundary injections. (e) The IR exponent of the distribution function of $R_3$ for bulk ($\tau_2$) and boundary ($\tau_2(z=0)$) injections in terms of $\alpha$. The corresponding confined area has been shown in the inset.}
\label{fig:3Dexponents}
\end{figure*}
Fig.~\ref{fig:M3_R3} shows the plot of $\left\langle \log(M_3)\right\rangle$ in terms of $\left\langle \log(R_3)\right\rangle$ whose slope is $\gamma_{M_3R_3}\equiv D_F^{M_3}$ which is the 3D mass fractal dimension for $L=300$ and various $\alpha$'s. We note that $D_F^{M_3}(\alpha=0)\simeq 2.96\pm 0.02$~\cite{dashti2015statistical}. Interestingly it is seen that the graphs smoothly cross over to the large scale regions in which the slope (fractal dimension) ($m_{\text{IR}}\equiv \gamma_{M_3R_3}^{\text{IR}}$) is different from the slope in the small-scale region with the slope $m_{\text{UV}}\equiv \gamma_{M_3R_3}^{\text{UV}}$. We name the small scales as \textit{UV limit}, and the large scales as \textit{IR limit}. The point of this change of behavior is $\alpha$-dependent. This point can easily be calculated using the linear fit of the graphs in each individual region. The transition point ($R_3^*$) is simply the point in which the fits meet each other. The fact that $m_{\text{UV}}(\alpha)$ is nearly $\alpha$-independent and $m_{\text{IR}}(\alpha)$ runs crucially by varying $\alpha$ can be seen in the upper inset of Fig.~\ref{fig:M3_R3}. We interpret $R_3^*$ as the point at which the cross-over takes place to the large scale properties since for $r\lesssim R_3^*$ the results are very close to the regular BTW model, whereas for $r\gtrsim R_3^*$ the behavior is different and not universal (presumably mixed with the finite-size effects). More interestingly we have observed that $R_3^*$ is a decreasing function of $\alpha$, i.e. $R_3^*\sim\alpha^{-\zeta}$ in which $\zeta=0.5\pm 0.05$. Since $\alpha$ can be interpreted as the measure of how directly a random chosen site is connected to a boundary site at which dissipation occurs, we can say that effectively (on average) a fraction of grains are dissipated in a bulk toppling depending on the amount of $\alpha$. The other effect is the \textit{sink} role of connection sites in micro-avalanches. In fact when $\alpha$ increases, the probability that a micro-avalanche involves a site which has a long-range link to the other micro-avalanches increases. Since, roughly speaking, such sites play the role of sink points, one may expect that the effective model for micro-avalanches be a dissipative one. It is known that the dissipative BTW model is equivalent to the massive ghost action $S=\int{d^{2}z}\left( \partial\theta\bar{\partial}\bar{\theta}+\frac{m^{2}}{4}\theta\bar{\theta}\right)$, where $\theta$ and $\bar{\theta}$ are complex Grassmann variables and $m^{2}$ is the number of sand grains dissipated in each toppling ($m$ can be fractional). On the other hand it is known that $R_3^*\sim m^{-1}$~\cite{Najafi2012Avalanche}. From these two points one concludes that effectively our model is equivalent to the dissipative BTW model with $m^2\sim\alpha$. This correspondence is acceptable only for $r\lesssim R_3^*$ and shows that the large scale regime is directly affected by the dissipations in the boundary sites and finite size effects. This result is reasonable since the amount of grain dissipation in a single component of an avalanche (the number of sand grains which are transferred out of that area) is proportional to the number of nodes with long-range links in that area. The problem description is not however as simple as stated above since there are surely some other links that return energy to the original micro-avalanche which partly compensates the dissipation effects. It is worth noting a comment concerning the numerical value of $\xi$ which is claimed to be $1/d\approx 0.33$ in three dimensions \cite{bhaumik2013critical} which is true for the total avalanches. For micro-avalanches however the statistics is different and it is acceptable that $R_3^*$ should be proportional to $m^{-1}\propto\alpha^{-1/2}$ which is representative of the grain dissipation towards the other micro-avalanches and is a well-known property of the dissipative sand-pile models. We have also considered the finite size effect of the results which have been shown in Figs. \ref{fig:M3_R3-alpha1}. The constant trend of $m_{\text{UV}}$ is seen in the inset of this figure, in which it is seen that its numerical value (for $\alpha=1$) is nearly robust against varying lattice size $L$, whereas $m_{\text{IR}}$ changes considerably by lattice size. This reflects the universal behavior of $m_{\text{UV}}$.\\
\begin{table*}[]
\begin{tabular}{|c|c|c|c|c|c|c|}
\hline & $M_3$ & $M_3(0)$ & $R_3$ & $R_3(0)$ & $n_{\text{toppling}}$ & $n_{\text{toppling}}(0)$\\
\hline $\tau(\alpha=0)$ & $1.34(4)$ & -- & $2.53(5)$ & -- & -- & -- \\
\hline $\tau_1$ & $1.37(4)$ & $1.6(2)$ & $2.07(5)$ & $2.8(5)$ & $1.33(1)$ & $1.66(2)$ \\ 
\hline $\tau_2(\alpha=1)$ & $3.46(5)$ & $3.62(9)$ & $5.98(7)$ & $6.1(7)$ & -- & -- \\ 
\hline $\gamma_{\tau_2}$ & $0.17(3)$ & $0.19(9)$ & $0.18(1)$ & $0.18(3)$ & -- & -- \\ 
\hline cut$(\alpha=1)$ & $2843$ & $4601$ & $11.2(5)$ & $13.9(4)$ & $3073$ & $6629$ \\
\hline $\gamma_{\text{cut}}$ & $1.28(9)$ & $1.23(5)$ & $0.42(8)$ & $0.49(7)$ & $1.57(3)$ & $1.06(3)$ \\
\hline $C(\alpha=1)$ & $9.6(6)$ & -- & $1.71(8)$ & -- & -- & -- \\
\hline $\gamma_C$ & $0.58(0)$ & -- & $0.156(5)$ & -- & -- & -- \\
\hline
\end{tabular}
\caption{The asymptotic values of the exponents. For each quantity there is a "cut" value in which a cross over between small scale behavior (which are consistent with regular 3D BTW model) and large non-universal behavior occur. It has been found these cut-values scale with $\alpha$ in a power-law fashion. For example $M_3^*\equiv M_3^{\text{cut}}=m^{\text{cut}}(\alpha=1)\alpha^{-\gamma_{\text{cut}}}$ which have been shown by "cut" in the table. In contrast to $\tau_1$, $\tau_2$ runs with $\alpha$ for all quantities; $\tau_2=\tau_2(\alpha=1)\alpha^{+\gamma_{\tau_2}}$ which have been shown separately in the table. The same is true for "confined area" abbreviated by $C$.} 
\label{tab:3d-tau}
\end{table*}
The same phenomena is seen in the distribution functions of $R_3$ and $M_3$. For $R_3$ (see Fig.~\ref{fig:P_R3}) as well as $M_3$ two distinct slopes are observed. The first slope is universal, which means that the slope of the first part of the graph is the same (for $L=300$, $\tau_r^{(1)}=2.1\pm 0.1$), whereas for the second part the slope ($\tau_r^{(2)}$) is non-universal and changes with $\alpha$ in the power law form, i.e. $\tau_r^{(2)}\sim \alpha^{\gamma_{\tau_2}}$ in which $\gamma_{\tau_2}=0.18\pm 0.05$ for $L=300$ and $0.1\leq \alpha\leq 10$. The transition point $R_3^*$ from UV to IR limits has also been shown in the inset of this figure. It is interestingly seen that $R_3^*\sim \alpha^{-\zeta_0}$ in which $\zeta_0=0.4\pm 0.1$ which is consistent with the result for $R_3^*$ in the Fig.~\ref{fig:M3_R3}. \\
It is worth mentioning that it is well-known in the literature that in the IR limit the BTW model on the random network phase (in which $\alpha\gtrsim 10$) should have the same exponents as the mean field (MF) results \cite{hoore2013critical,bhaumik2013critical,moosavi2015structural,bhaumik2016dissipative}. Two points should be noted in this regard: Firstly we have not entered this phase. In our analysis $\alpha$ has been considered up to $10$. Secondly, as stated above, our statistical observables (micro-avalanches) are different from the ones considered in the above mentioned references (total avalanches) and the statistics of micro-avalanches (which are claimed to act like the dissipative avalanches) is apparently different from the total avalanche as a whole. As stated above the site which connects the original micro-avalanche to the other ones plays (partly) the role of a sink point in this avalanche, leading it to behave like a dissipative avalanche with the mass $m^2\sim\alpha$ in its effective ghost action (dissipative BTW model). The fact that the exponents of the micro-avalanches are different from (higher than) the exponents for the total avalanches can be inferred from the following argument: Suppose that we are determining the exponent of the distribution function of the avalanche size. This distribution function falls off more rapidly than the distribution function of the total avalanche. As a result the exponent of the micro-avalanches is higher than the corresponding exponent of the total avalanche. It is notable that the above-mentioned effects appear in the IR-limit, i.e. for small scales the properties of the regular BTW model should be seen.  As the system enters the IR-regime, or equivalently the concentration of the long-range links increases, the effect of the internal structure of the model and the grain dissipation in the boundaries inevitably appear. To understand this, note that the length distribution of the long-range links is uniform (limited to the lattice size), i.e. the connection probability of a typical site to a bulk site is the same as the one for the boundary sites. This causes the system to experience simultaneously the effects of the IR limit (dissipative BTW model) and the dissipation from the boundaries. Due to this fact we see that the exponents in the large-scale limit are $\alpha$-dependent. \\ 

As stated above we have performed two parallel simulations: avalanches in the bulk with $z=L/2$ cross sections, and avalanches on the boundaries with injection point and cross sections on the $z=0$ plane. We name the first ones as the bulk avalanches and the second ones as the boundary avalanches. One of our observations is that as $\alpha$ increases, the form of the distribution functions and fractal dimensions of the bulk avalanches become more and more closer to the ones for the boundary avalanches. To quantify this trend, we define a measure which states how far the curves are, see Fig.~\ref{fig:confined}. In this figure two distribution functions for the same amount of $\alpha$ have schematically been shown: one for the bulk avalanche and the other for the boundary avalanche. Let us name the area between these two curves as $A$ and define $C_x(\alpha)\equiv\exp[A]$ which is numerically calculated as follows: Consider the mentioned two distribution functions of the $x$ variable; one for the bulk injection $N(x)$ and another for the boundary injection $N_0(x)$ and define $Y\equiv \ln N$ and $X\equiv \ln x$. Using the discrete integration relation for $A$ we find that $C_x(\alpha)$ is as follows:
\begin{equation}
C_x(\alpha)=\exp\left[\sum_{i=1}^{N}\delta Y\left(X_{\text{bulk}}^i-X_{\text{bdry}}^i\right)\right] =\prod_{i=1}^{N}\left(\frac{x_{\text{bulk}}^i}{x_{\text{bdry}}^i}\right)^{\delta Y}
\label{Eq.Cs}
\end{equation} 
in which we have divided the $Y$ axis (in the total range $[Y_{\text{min}},Y_{\text{max}}]$) to $N\equiv\frac{Y_\text{max}-Y_\text{min}}{{\delta Y}}$ sub-intervals with the lengths $\delta Y$ (= constant) and $X_{\text{bulk}}\equiv\ln(x_{\text{bulk}})$ and $X_{\text{bdry}}\equiv\ln(x_{\text{bdry}})$ have been defined in the Fig.~\ref{fig:confined}. $Y_{\text{max}}$ is defined as the point at which the curves meet each other for the first time (small $X$s) and $Y_{\text{min}}$ is the minimum point in our analysis. For the case in which the graphs coincide, we have $C_x(\alpha)=1$ resulting to $A=0$ as expected. The smaller the numerical amounts of $C$'s are, the closer the bulk statistics are to the boundary statistics. This definition may seem extravagant, since $Y_{\text{min}}$ can clearly be decreased unboundedly which increases $A$. In fact this is not annoying, since decreasing $Y_{\text{min}}$ does not alter the exponents of $C_x$'s (which are reported bellow) for small enough $Y_{\text{min}}$'s, since both graphs become nearly vertical and the fractions in the Eq. \ref{Eq.Cs} generate some multiplicative number which increases by decreasing $Y_{\text{min}}$, i.e. the exponent is not altered. Therefore the definition Eq. \ref{Eq.Cs} is well-defined. The other point is that $A$ becomes smaller for larger $\alpha$'s which shows that the boundary (dissipative) sites become more accessible for bulk sites for larger $\alpha$ in such a way that the system becomes more dissipative. The interesting feature is that for all $x$'s, $C_x(\alpha)$ has a nice power-law behavior in terms of $\alpha$ in the interval of interest, i.e. $C_x(\alpha)=C_x(\alpha=1)\alpha^{-\gamma_{C_x}}$. The benefit of the chosen form of $C_x$ in Eq. \ref{Eq.Cs} is that it shows (on average) the $\frac{x_{\text{bulk}}}{x_{\text{bdry}}}$ ratios show power-law behaviors with $\alpha$. The numerical amounts for $C_x(\alpha=1)$ and $\gamma_{C_x}$ have been presented in the TABLE~\ref{tab:3d-tau} from which we see that $\gamma_{C_{M_3}}=0.58\pm 0.01$ and $\gamma_{C_{R_3}}=0.16\pm 0.01$ for $L=300$ and the $\alpha$ interval of interest.\\
The exponents of the second part of the graphs are more interesting, since they presumably carry the effect of the boundary sites and the internal structure of the model. The non-universal character of these quantities can be distinguished in Fig.~\ref{fig:P_R3}. Our observation is that it has power-law behavior in terms of $\alpha$, i.e. $\tau_2=\tau_2(\alpha=1)\alpha^{\gamma_{\tau_2}}$ for which the exponents $\gamma_{\tau_2}$ have been reported separately in TABLE~\ref{tab:3d-tau} for $L=300$. In the Fig.~\ref{fig:tau2_alpha}, $\tau_2$ along with the confined area for the bulk and boundary avalanches have been shown for 3D gyration radius. The most straight and nice power-law behavior in these systems is seen for the \textit{local} quantities, i.e. the number of toppling in an avalanche $N(n_{\text{toppling}})$. A cross over has been observed for this quantity at some $n$ value, i.e. $n^*$ below which the exponent $\tau_N$ is identical for all values of $\alpha$. $n^*$ has also power-law behavior in terms of $\alpha$ which has been reported in TABLE~\ref{tab:3d-tau}.\\
We see that there are two scales with different physics: for the UV limit the system behaves just like ordinary regular BTW model, whereas for the IR limit the system shows non-universal critical behavior which is most consistent with the dissipative BTW model. For larger $\alpha$'s the large-scale phase becomes more dominant showing that $\alpha$ favors the large-scale (dissipative) behaviors which is expected.\\
A point concerning the concreteness of the numerical evaluation of the cross-over points should be mentioned here. In obtaining the cross-over points, it should be noted that that in a graph with two distinct linear fitting, one should pay especial attention to the numerical error bar of the cross point (in which the linear fits meet each other).  Let us suppose that two linear fits are: $y_1=\alpha_1x+\beta_1$ and $y_2=\alpha_2x+\beta_2$. The relative error for the transition point is simply calculated to be $\left( \delta x^*/x^*\right)^2=\left(\delta \alpha_1^2+\delta \alpha_2^2\right)/\left(\alpha_1-\alpha_2\right)^2+\left(\delta \beta_1^2+\delta \beta_2^2\right)/\left(\beta_1-\beta_2\right)^2$ in which $\delta \beta_i$ and $\delta \alpha_i$ are the errors of $\alpha_i$ and $\beta_i$. . In the cases in which $\alpha_1$ and $\alpha_2$ or $\beta_1$ and $\beta_2$ are close to each other, $\delta x^*/x^*$ becomes large, leading to unreliable results. Fortunately in our work none of $\alpha$'s nor $\beta$'s are close to each other. For instance, in Fig. \ref{fig:M3_R3} for $L=300$ and $\alpha=0.5$, $\delta x^*/x^*\approx 0.24$, for Fig.\ref{fig:P_R3} it is $\delta x^*/x^*\approx 0.1$, for Fig. \ref{fig:P_M2} it is $\delta x^*/x^*\approx 0.06$ and for Fig. \ref{fig:P_r} it is $\delta x^*/x^*\approx 0.04$. We see that except for the first case the relative errors are reasonably small. The relative closeness of the slopes in the first case is the reason of this fact.

\section{2D Induced model}
\label{sec:2D}

As stated above we consider two cross sections: one at $z=0$ for boundary avalanches and the other at $z=L/2$ for bulk avalanches. We investigate the statistical quantities $x=a,M_2,l,r,R_2$ and $n_{\text{toppling}}$ which are statistical observables for two-dimensional cross sections of micro-avalanches (2DCSMA), defined in SEC~\ref{sec:ASM}. All of these quantities have also been calculated for the boundary avalanches. As before we have two types of the exponents: the fractal dimensions which are the exponents $\gamma_{xy}$ in the relations $\left\langle \log(y)\right\rangle=\gamma_{xy} \left\langle \log(x)\right\rangle +\text{cst.}$, and the exponents of the distribution functions $\tau_x$ defined by the relation $N(x)\sim x^{-\tau_x}$. Just like the previous section, we have observed that the system behaves like the regular BTW model for the UV limit (small scales), whereas for the IR limit (large scales) the behaviors are changed. The other observation is that the graphs of the bulk 2DCSMAs tend to behave more like the boundary 2DCSMAs for larger $\alpha$'s. To measure this, we have calculated the confined area for the distribution functions of all quantities like the previous section and observed a power-law behavior in terms of $\alpha$.

\subsection{Distribution Functions}
\label{sec:2D-distribution}

\begin{figure*}
\centering
\begin{subfigure}{0.45\textwidth}\includegraphics[width=\textwidth]{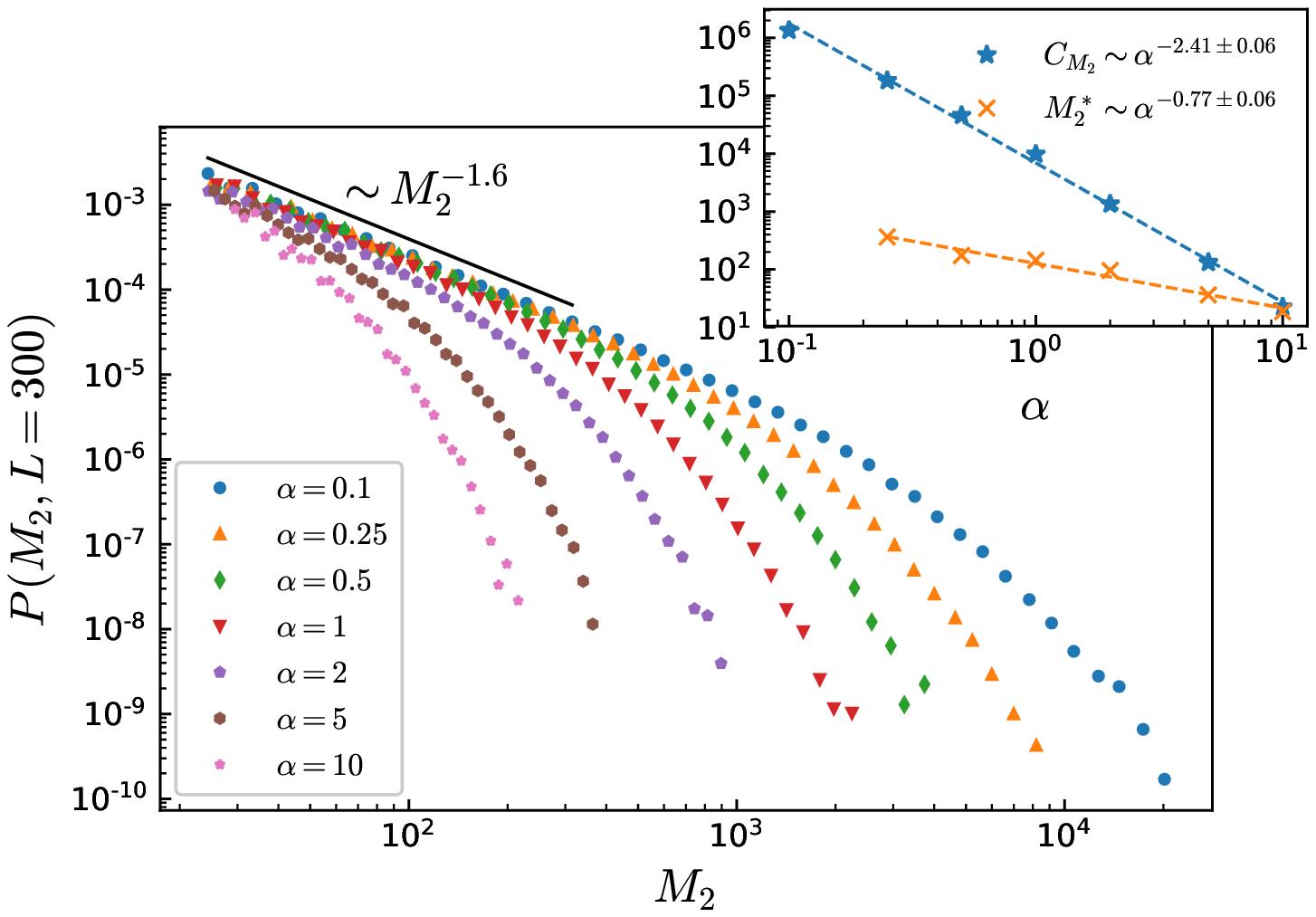}
\caption{}
\label{fig:P_M2}
\end{subfigure}
\begin{subfigure}{0.45\textwidth}\includegraphics[width=\textwidth]{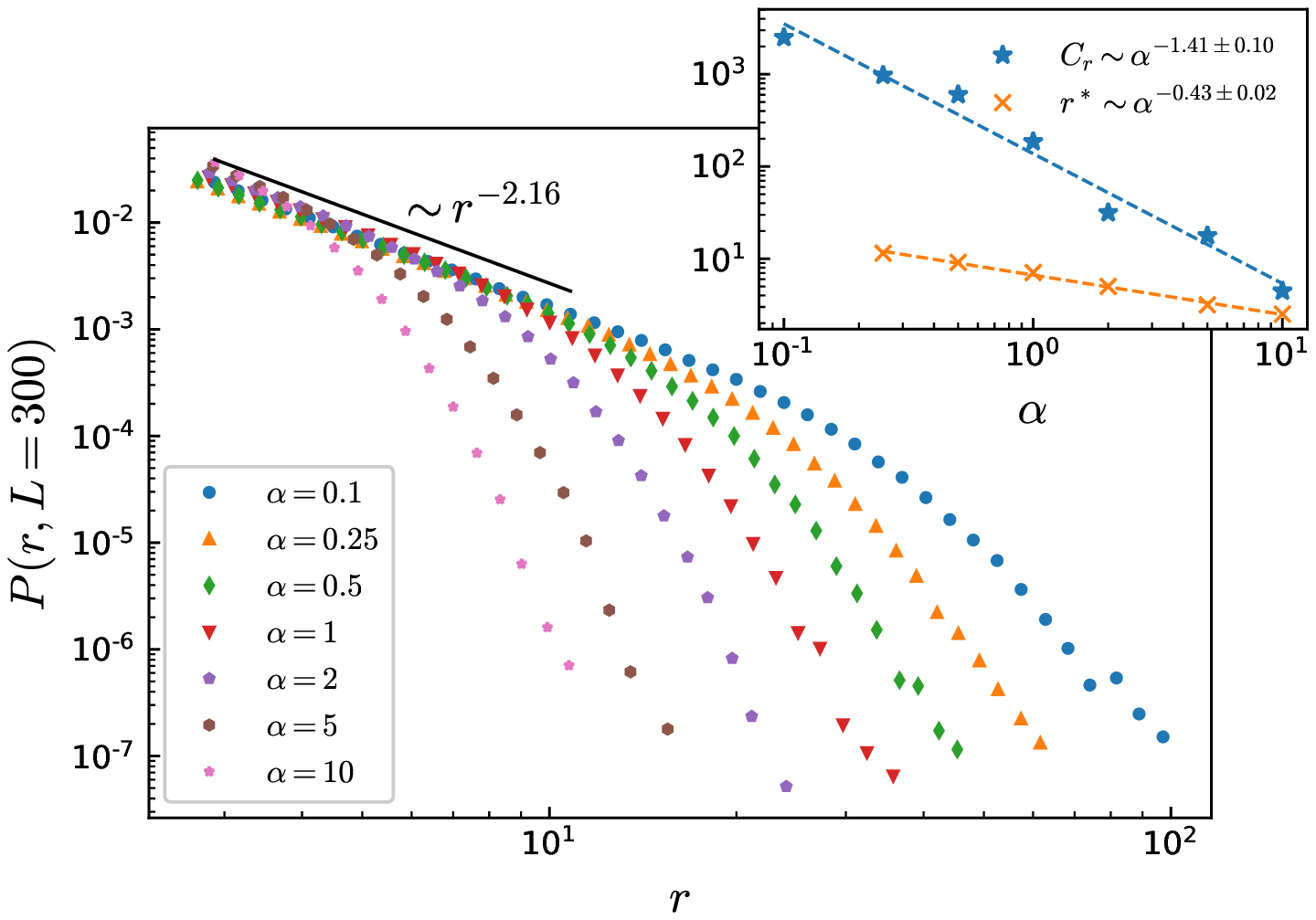}
\caption{}
\label{fig:P_r}
\end{subfigure}
\caption{The log-log plot of the distribution function of (a) $M_2$ and (b) $r$ for various rates of $\alpha$. Insets: the power-law behavior of the confined area of (a) $C_{M_2}$ and $M_2^*$ and (b) $C_r$ and $r^*$ in terms of $\alpha$.}
\label{fig:2D-loop-distro}
\end{figure*}

The power-law behavior, expected for critical systems is observed for the distribution functions, as is seen in the Fig.\ref{fig:2D-loop-distro}. We have found that the graphs approach to the boundary graphs in the power-law fashion in terms of $\alpha$ which has been quantified by $C_{x}(\alpha)$. In the Fig. \ref{fig:P_M2} the log-log plot of the mass distribution function of two-dimensional clusters has been shown for various rates of $\alpha$. The UV slope of the graph is nearly $1.6$ for $L=300$, whereas for IR limit the slope is non-universal which can be interpreted as the fingerprint of finite size effects. The power-law behavior of $C_{M_2}(\alpha)$ has been shown in the inset of this figure in which it is shown that $C_{M_2}\sim \alpha^{-\gamma_{C_{M2}}}$ and $M_2^*\sim \alpha^{-\gamma_{M_2^{\text{cut}}}}$ in which $\gamma_{C_{M2}}\simeq 2.4$ and $\gamma_{M_{2}^{\text{cut}}}\simeq 0.77$ for $L=300$. Let us explain this result in terms of the fractal dimension $M_2\sim R_2^{\gamma_{M_2R_2}}$ in which, as we will see in the next section $\gamma_{M_2R_2}\simeq 1.88$ for low $\alpha$ values. From this equation we easily see that $M_2^*\sim (R_2^*)^{\gamma_{M_2R_2}}$. From the TABLE\ref{tab:3d-tau} we know that $R_2^*\sim R_3^*\sim \alpha^{0.42}$ from which we obtain $M_2^*\sim \alpha^{-0.78}$ which is consistent with the above result. The same log-log plot has been drawn for the loop gyration radius $r$ in the Fig. \ref{fig:P_r}. The approach to the boundary curves with the exponent $1.4$ (for $L=300$) is evident. The cut value of radius $r^*\simeq \alpha^{0.43}$ coincides with the exponent for $R_3^*$ which shows that the approximate duality to the massive ghost model is also preserved for the cross-sections. The full information of the exponents of the distribution functions of the statistical observables have been gathered in TABLE\ref{tab:2d-tau}. In this table the UV exponents have been shown by $\tau_1$ which is nearly independent of $\alpha$. The IR exponents $\tau_2$ depend on $\alpha$ in a power-law fashion in the considered interval, i.e. $\tau_2(\alpha)=\tau_2(\alpha=1)\alpha^{\gamma_{\tau_2}}$. The cut-values also show power-law behaviors in terms of $\alpha$. It is worth mentioning that there are some hyper-scaling relations between these exponents as stated above. Interestingly we have $\gamma_{C_{M_2}}\simeq \gamma_{C_{R_2}}\times\gamma_{M_2,R_2}$ which cannot be explained in terms of simple scaling relations. The same is also true for other $\gamma$'s. \\
We have calculated the UV exponents ($\tau_1$) for various rates of lattice sizes and have seen that all of the exponents are nearly saturated for maximum lattice size in this work, i.e. $1/L=0.0033$, showing that the results for $L=300$ are reliable.
\begin{table*}[]
\begin{tabular}{|c|c|c|c|c|c|c|c|}
\hline & $a$ & $l$ & $r$ & $R_2$ & $M_2$ & $n_{\text{toppling}}$ & $n_{\text{toppling}}(0)$\\
\hline $\tau(\alpha=0)$ & $1.63(4)$ & $1.88(4)$ & $2.21(5)$ & -- & $1.58(4)$ & -- & -- \\
\hline $\tau_1$ & $1.6(1)$ & $1.9(2)$ & $2.2(5)$ & $2.05(5)$ & $1.58(1)$ & $1.0(2)$ & $0.98(1)$ \\ 
\hline $\tau_2(\alpha=1)$ & $5.5(3)$ & $7.7(8)$ & $10.1(1)$ & $9.3(1)$ & $7.1$ & -- & -- \\ 
\hline $\gamma_{\tau_2}$ & $0.11(9)$ & $0.2(2)$ & $0.18(5)$ & $0.17(5)$ & $0.2(1)$ & -- & -- \\ 
\hline cut$(\alpha=1)$ & $149$ & $110$ & $6.7(5)$ & $5.3(1)$ & $127$ & $131$ & $190$ \\
\hline $\gamma_{\text{cut}}$ & $0.77(3)$ & $0.45(3)$ & $0.43(2)$ & $0.45(4)$ & $0.77(5)$ & $0.9(2)$ & $0.68(1)$ \\
\hline $C(\alpha=1)$ & $4.32(5)$ & $2.4(1)$ & $2.1(4)$ & $1.64$ & $3.8(1)$ & -- & -- \\
\hline $\gamma_C$ & $2.5(4)$ & $1.48(4)$ & $1.41(1)$ & $1.38(1)$ & $2.41(5)$ & -- & -- \\
\hline
\end{tabular}
\caption{The asymptotic values of the exponents. The symbols are the same as TABLE\ref{tab:3d-tau}.} 
\label{tab:2d-tau}
\end{table*}

\subsection{Fractal ِDimensions}
\label{sec:2D-FDs}
\begin{figure*}
\centering
\begin{subfigure}{0.45\textwidth}\includegraphics[width=\textwidth]{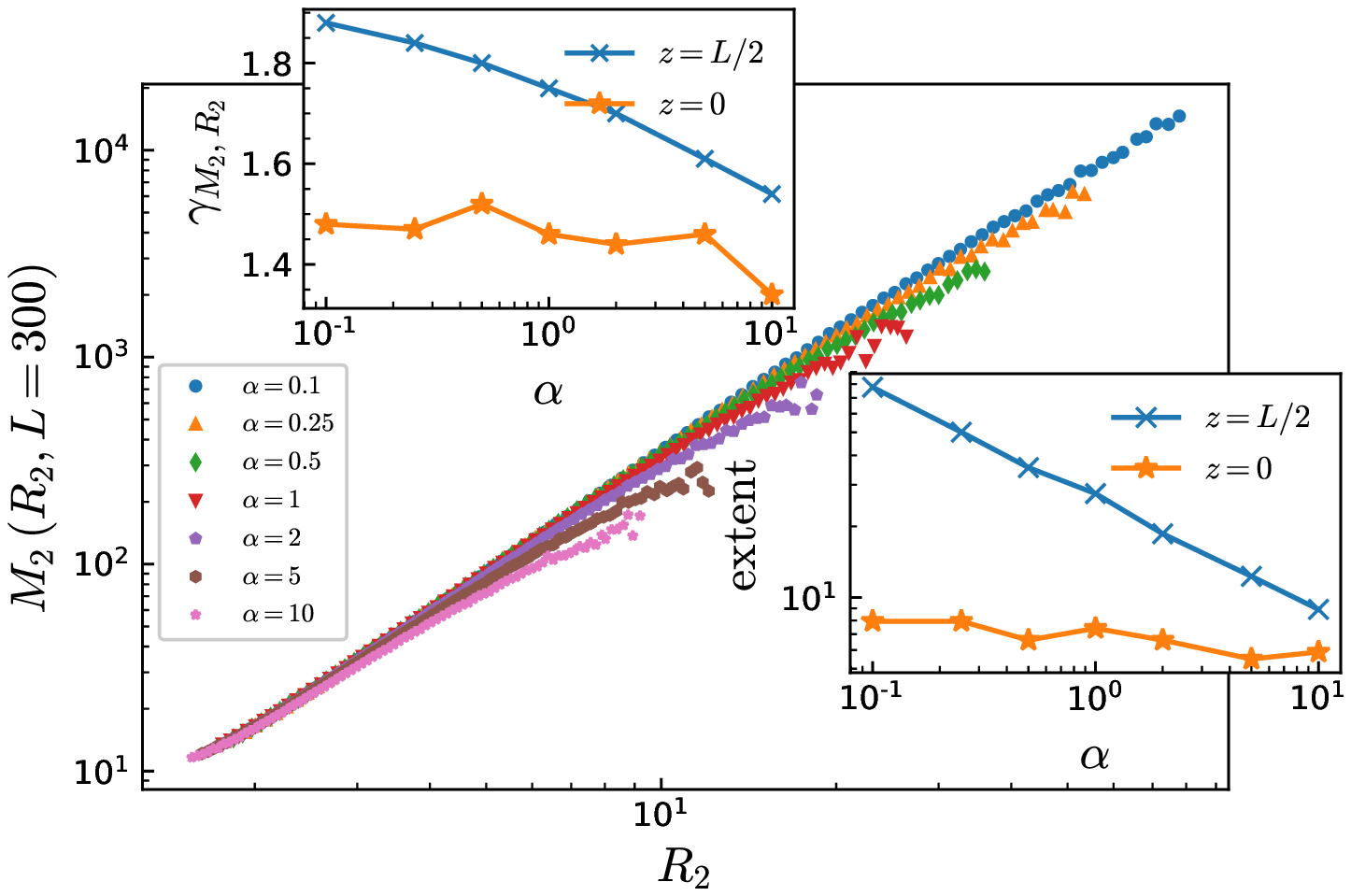}
\caption{}
\label{fig:M2_R2}
\end{subfigure}
\begin{subfigure}{0.45\textwidth}\includegraphics[width=\textwidth]{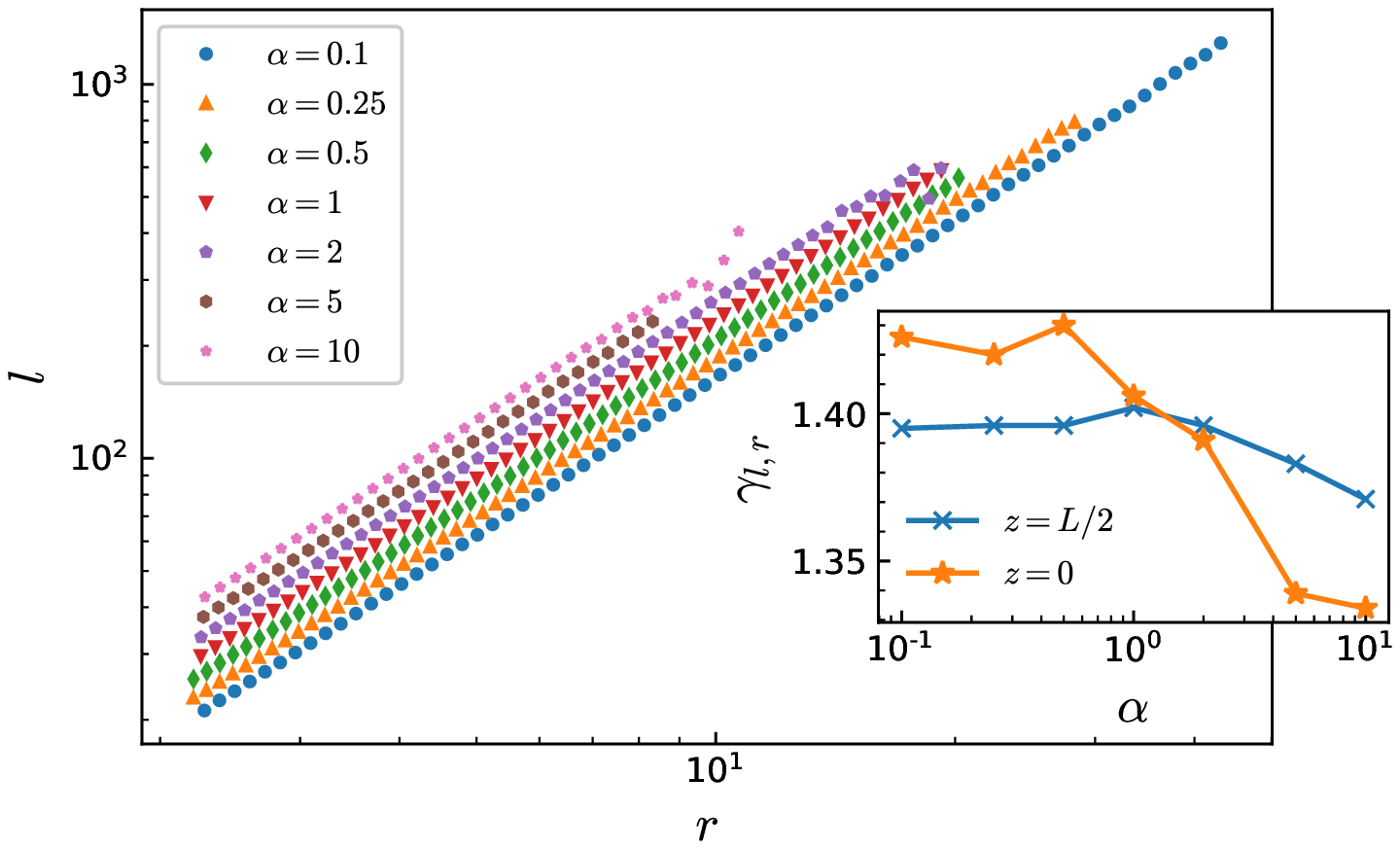}
	\caption{}
	\label{fig:l_r}
\end{subfigure}
\begin{subfigure}{0.45\textwidth}\includegraphics[width=\textwidth]{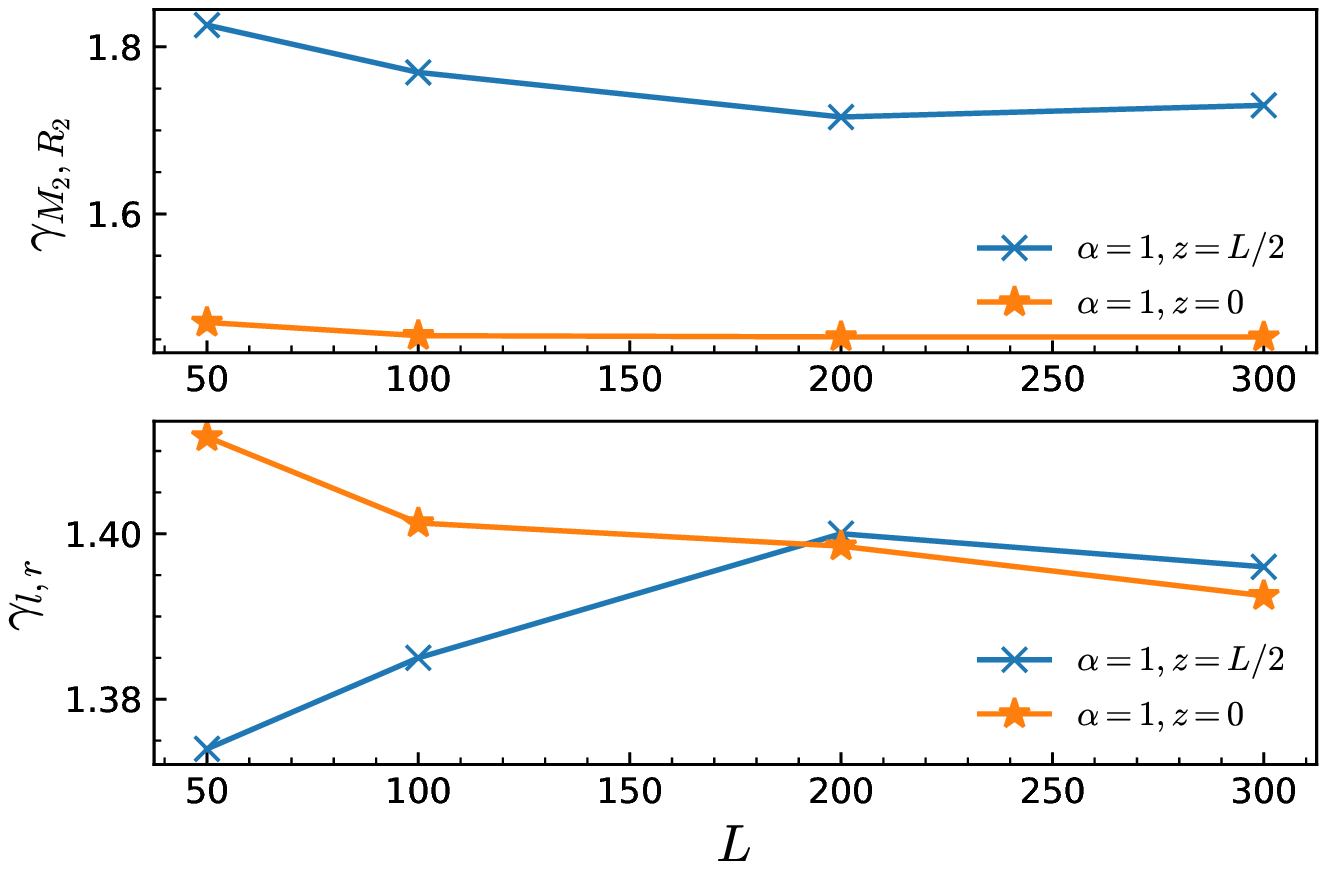}
\caption{}
\label{fig:gama_M2R2_L}
\end{subfigure}
\begin{subfigure}{0.45\textwidth}\includegraphics[width=\textwidth]{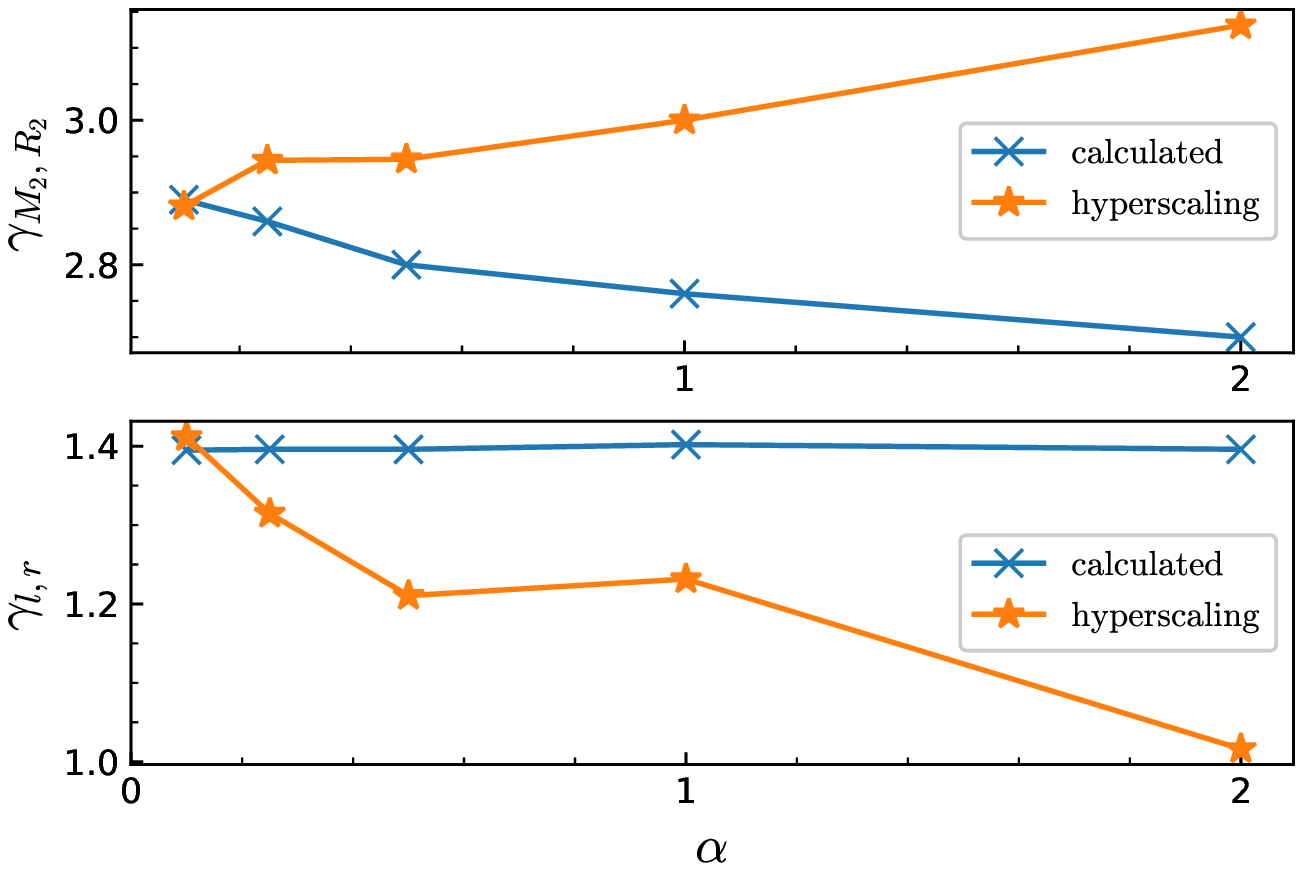}
\caption{}
\label{fig:hyper_calc}
\end{subfigure}
\caption{(a) The fractal dimension $\gamma_{M_2R_2}$. Upper inset: the exponents for $z=L/2$ and $z=0$ cross-sections. Lower inset: spatial extents of regular BTW behavior for $z=L/2$ and $z=0$ cross-sections. (b) The fractal dimension $\gamma_{lr}$. Inset: The exponents for $z=L/2$ and $z=0$ cross-sections in terms of $\alpha$. (c) The finite size dependence of the exponents. (d) The hyper-scaling exponents $\gamma_{x,y}\equiv\frac{\tau_y-1}{\tau_x-1}$ and the calculated exponents in terms of $\alpha$.}
\label{fig:2D-Geo}
\end{figure*}

The fractal dimensions are very powerful tools for identifying the universality class of any critical model. In this section various fractal dimensions ($\gamma_{xy}$'s defined in the previous section) for cross sections are processed. A very smooth change of behavior from small scales to large scales is seen for these quantities. For example in Fig. \ref{fig:2D-Geo} two fractal dimensions ($\gamma_{M_2R_2}$ and $\gamma_{l,r}$) have been shown and analyzed.
\begin{table*}
\begin{tabular}{|c|c|c|c|c|c|c|}
\hline & $\gamma_{la}$ & $\gamma_{la}(0)$ & $\gamma_{lr}$ & $\gamma_{lr}(0)$ & $\gamma_{M_2R_2}$ & $\gamma_{M_2R_2}(0)$ \\ 
\hline $\gamma(\alpha=1)$ & $0.7(1)$ & $0.7(5)$ & $1.395(5)$ & $1.392(5)$ & $1.73(5)$ & $1.45(3)$ \\ 
\hline $\tau_{\gamma}$ & $0.005(1)$ & $0.02(1)$ & $0.0$ & $0.016(1)$ & $0.044(5)$ & $0.017(5)$ \\ 
\hline $\gamma_{xy}|_{\alpha=1}=\frac{\tau_y-1}{\tau_x-1}$ & $0.67(3)$ & - & $1.2(3)$ & - & $1.6(4)$ & -\\
\hline
\end{tabular}
\caption{The calculated fractal dimensions $\gamma_{x,y}$ with the relation $\gamma_{x,y}\sim \gamma_{x,y}(\alpha=1)\times \alpha^{-\tau_{\gamma}}$ for both $z=L/2$ (without argument) and $z=0$ (with argument $(0)$).} 
\label{tab:2d-gamma}
\end{table*}
In Fig. \ref{fig:M2_R2} the mentioned change of behavior for $\gamma_{M_2R_2}$ ($L=300$) is obvious. The deviation from the UV (small scale) slopes are $\alpha$-dependent. This cross over is so smooth that a distinct point for which the slopes of the graph above and below it are sharply different cannot be singled out. Instead we have calculated the mean slope of the graph over the full interval which is shown in the upper inset of this figure in terms of $\alpha$. Noting that $\gamma_{M_2R_2}^{\alpha=0}\simeq 1.99$~\cite{dashti2015statistical}, we see that the extrapolation of the graph for $\alpha\rightarrow 0$ for $z_{\text{cross-section}}=L/2$ is compatible with the other works. For $z=0$-2DCSMAs $\gamma_{M_2R_2}^{\alpha=0.1}\simeq 1.45\pm 0.02$ and is nearly constant over the interested $\alpha$ interval. This, along with the other results for the boundary 2DCSMAs (see \ref{fig:P_r}), show that the properties of the model on the boundary plate \textit{are nearly $\alpha$-independent}. It is expected since, as stated in the previous sections, $\alpha$ is directly connected to the fact how bulk sites are connected to the boundary sites and for larger $\alpha$'s the effects of the boundary dissipations are more evident. However for boundary plate avalanches the effect of the dissipation is maximal. The very small dependence to $\alpha$ for these avalanches is the effect of the boundary sites on the other sides of the system which is negligible. The maximum spatial gyration radius in these graphs has been sketched in terms of $\alpha$ in the lower inset from which again the exponent $\simeq \frac{1}{2}$ is evident for bulk 2DCSMAs, just like the three-dimensional avalanches in which $\zeta\simeq \frac{1}{2}$ (see SEC.\ref{sec:3D}). This shows that the effective model in bulk 2D cross-sections is also massive with $m^2\sim \alpha$. Note that this function for the boundary 2DCSMAs is nearly constant for all $\alpha$s. The same graphs have been sketched for $\gamma_{l,r}$ in Fig. \ref{fig:l_r} whose inset shows this exponent in terms of $\alpha$ for $z_{\text{cross-section}}=0,L/2$. For $\alpha=0.1$ and $L=300$ one retrieves nearly the $\alpha=0$ result, i.e. $\gamma_{l,r}(\alpha=0)\equiv D_F(\alpha=0)\simeq \frac{11}{8}=D_F^{\text{Ising}}$. This result is nearly independent of $\alpha$ for bulk 2DSCMAs, but for boundary 2DCSMAs it is $\alpha$-dependent and approaches $\simeq 1.43$ as $\alpha\rightarrow 0.1$ for $L=300$. The finite size effects of these exponents have been shown in Fig. \ref{fig:gama_M2R2_L} which show saturation in $L=300$. \\
A very important check is a hyper-scaling relation between $\tau$'s and $\gamma$'s, namely $\gamma_{xy}=\frac{\tau_y-1}{\tau_x-1}$ \cite{jensen19891,lubeck2000moment,najafi2015geometrical}. This relation is valid only when the conditional probability $P(x|y)$ is a very narrow function of both $x$ and $y$. We have observed that this hyper-scaling relation holds only for the low $\alpha$'s, i.e. for large and medium values this relation is violated meaning that $P(x|y)$ is not a narrow function in this limit. These are shown in Fig. \ref{fig:hyper_calc} in which it is seen that the graphs become more and more separated for large $\alpha$'s. The total data about the exponents and their variation in terms of $\alpha$ have been shown in TABLE\ref{tab:2d-gamma}.

\section{Conclusion}
\label{Conclusion0}

This paper has been devoted to the problem of the energy propagation in 3D small world networks and their 2D cross-sections with the long-rang link fraction $\alpha$. For three-dimensional case, as well as two dimensional one, a smooth cross over from regular BTW (UV) limit to purely dissipative (corresponding to a massive ghost action) one has been observed. By analyzing the spatial extent up to which the regular BTW model was observed, it was revealed that for the large scale (IR) limit the system behaves like a dissipative BTW model with the dissipation factor (the mass in the corresponding ghost action) $m^2\sim\alpha$. This result may be expected since $\alpha$ is a parameter representing how short the mean least path between a typical bulk site and boundary site (at which energy dissipation occurs) is.\\
Our reason for analyzing 2D cross section was twofold: Firstly for many experiments the array of energy activity detectors covers only partly the nodes of the system, which may be considered for example a 2D lattice. Secondly from the theoretical side, our motivation was the question how the information in $d+1$ system spread in the $d$-dimensional slices. The induced model on $z_{\text{cross-section}}=L/2$ is critical with exponents distinct from the 3D ones compatible with previous results which have been reported in the paper. The spatial extents, up to which the BTW-type critical behaviors are seen, are $\alpha$-dependent and for extreme dissipative (large $\alpha$) limit the power-law behaviors disappear. For boundary cross-section however, the critical extent is very low showing its extreme dissipative character.\\
Along with the cross-section in $z=L/2$, we have also analyzed the energy spread in $z_{\text{cross-section}}=0$ planes to observe how the induced model on $z=L/2$ cross-sections approaches the induced model on $z_{\text{cross-section}}=0$ planes as a function of $\alpha$ by defining a measure, i.e. $C_x(\alpha)$. Interestingly it was revealed that all statistical observables on $z_{\text{cross-section}}=L/2$ cross-sections approach to the boundary counterparts ($z_{\text{cross-section}}=0$ cross-sections) in a power-law fashion. The exponent of these behaviors is observable-dependent which has been reported in the paper. For both 3D and 2D models the behavior of all of the quantities can be divided to two scales: for small (UV) scales the exponents are $\alpha$-independent and are compatible with regular BTW model, whereas for large (IR) scales they are power-law $\alpha$-dependent. The exponents of these relationships are $L$(=lattice size)-dependent.\\
The hyper-scaling relation between the exponents of distribution functions and fractal dimensions is another important issue which has been addressed in this paper. We numerically showed that they are satisfied for small $\alpha$s and are violated for large $\alpha$s.

\bibliography{refs}

\end{document}